\documentclass{aa}  

\usepackage[breaklinks=true,colorlinks,citecolor=blue]{hyperref}
\usepackage{graphicx}
\usepackage{enumerate}
\usepackage{epsfig}
\usepackage{color}
\usepackage{txfonts}
\usepackage{natbib}
\usepackage{multirow}
\usepackage{ulem} 

\definecolor{ao}{rgb}{0.0, 0.5, 0.0}

\newcommand{\Msun}{M\ensuremath{_\odot}}

\begin{document} 

\title{Stellar metallicity variations across spiral arms in disk galaxies with multiple populations}
\titlerunning{Stellar metallicity variations across spiral arms in disk galaxies with multiple populations}

\author{S. Khoperskov\inst{1}\thanks{sergey.khoperskov@obspm.fr}, P. Di Matteo\inst{1}, M. Haywood\inst{1}, F. Combes\inst{2,3}}
\authorrunning{S. Khoperskov et al.}

\institute{GEPI, Observatoire de Paris, PSL Universit{\'e}, CNRS,  5 Place Jules Janssen, 92190 Meudon, France
 \and Observatoire de Paris, LERMA, CNRS, PSL Univ., UPMC, Sorbonne Univ., F-75014, Paris, France \and College de France, 11 Place Marcelin Berthelot, 75005, Paris, France}

\date{Received ; accepted }
 
\abstract{This letter studies the formation of azimuthal metallicity variations in the disks of spiral galaxies in the absence of initial radial metallicity gradients. Using high-resolution $N$-body simulations, we model composite stellar discs, made of kinematically cold and hot stellar populations, and study their response to spiral arm perturbations. We find that, as expected, disk populations with different kinematics respond differently to a spiral perturbation, with the tendency for dynamically cooler populations to show a larger fractional contribution to spiral arms than dynamically hotter populations. By assuming a relation between kinematics and metallicity, namely the hotter the population, the more metal-poor it is, this differential response to the spiral arm perturbations naturally leads to azimuthal variations in the mean metallicity of stars in the simulated disk. Thus, azimuthal variations in the mean metallicity of stars across a spiral galaxy are not necessarily a consequence of the reshaping, by radial migration, of an initial radial metallicity gradient.  They indeed arise naturally also in stellar disks which have initially only a negative vertical metallicity gradient.}

   \keywords{galaxies: evolution --
             galaxies: kinematics and dynamics --
             galaxies: structure}

\maketitle

\section{Introduction}\label{sec::intro}
The physical origin of spiral arms in disk galaxies is a classical problem of galactic astronomy~\citep{2016ARA&A..54..667S}. Spiral structures can be interpreted as stationary discrete density waves~\citep[][and references therein]{1964ApJ...140..646L,1977PNAS...74.4726B,1978ApJ...226..508L,1996ssgd.book.....B}.  Some studies, using $N$-body simulations, have pointed out that spiral arms in pure stellar disks are short-lived structures~\citep[e.g.,][]{2011MNRAS.410.1637S,2013ApJ...763...46B}, but the increase of numerical resolution~\citep[e.g.,][]{2011ApJ...730..109F,2013ApJ...766...34D} or the modeling of multicomponent stellar-gaseous disks~\citep{2002A&A...384..872O,2012MNRAS.427.1983K,2015MNRAS.451.1350G} have demonstrated that spiral waves can maintain their morphological structure for a longer time. Independently of their nature, spiral structures are seen in all  disk components of  late-type galaxies~\citep[see, e.g.,][]{1995ApJ...447...82R,2002ApJS..143...73E} where they appear as azimuthal variations in the baryonic density distribution.

The presence of spiral arms in stellar component can stimulate the formation of molecular clouds~\citep[e.g.,][]{2006MNRAS.365...37B} and enhance star formation~\citep[e.g.,][]{2008AJ....136.2846B}. Spirals can also induce radial migration of stars~\citep{2002MNRAS.336..785S}. Hence their presence in a galaxy disk can  generate differences in the mean chemical composition  {of stars in} the arm and inter-arm regions. The existence of azimuthal variations in the chemical composition  {of stars has been} well established by observations of different external spiral galaxies~\citep[e.g.,]{2009ApJ...696.2014D, 2013ApJ...766...17L, 2017ApJ...846...39H,2017A&A...601A..61V,2017arXiv171001188S}. In the Milky Way disk,  {the} detection of azimuthal metallicity patterns  {has} become possible quite recently,  {thanks to} large spectroscopic surveys. For instance, analysing RAVE data, \cite{2017A&A...601A..59A} detected asymmetric metallicity patterns for a sample of stars in a cylinder of $0.5$~kpc radius from the Sun. Using  APOGEE red-clump stars, \cite{2014ApJ...790..127B}  {quantified the strength of} azimuthal variations  {in} the median metallicity  {of stars at few kpc from the Sun} to be of the order of ~$0.02$~dex. \cite{2009ApJ...696.2014D} reported  strong~($0.4$~dex) azimuthal gradients for O, Mg, Si combining data from HII regions, Cepheids, B-stars and Red Supergiant stars in the inner disk  {in a region possibly associated to} the end of the Milky Way bar. These trends are  {usually} interpreted as  {due to the} orbital effects of the Galactic bar and  {to the presence of} radial metallicity gradients in the disk. 

Despite recent observational progress, only a few theoretical studies have investigated azimuthal variations  {in the} metallicity~(or abundances)  {of stars in a disk galaxy and their link with} the presence of spiral structures.  {It has been shown that azimuthal variations can be generated by radial migration in a stellar disk which has initially a negative radial metallicity gradient.}
For instance, \cite{2013A&A...553A.102D} have shown that radial migration induced by a bar leads to significant azimuthal variations in the metallicity distribution of old stars. The combination of both tangential and radial velocity asymmetries creates streaming motions along spirals, with metal-rich stars from the inner disk that migrate through spiral patterns to the outer parts of the disk~\citep[see, e.g.,][]{2012A&A...548A.126M,2013MNRAS.436.1479K,2014MNRAS.443.2757K,2015MNRAS.453.1867G}.  Recently \cite{2016MNRAS.460L..94G} used a high-resolution cosmological zoom simulation of a Milky Way-sized halo to demonstrate that radial flows associated with  spiral arm s can produce an overdensity of metal-rich stars on the trailing side of the spiral  and an overdensity of metal-poor star on the leading side. Thus, in all these works azimuthal metallicity variations are the result of  {the} transformation of  {a} pre-existing radial negative metallicity gradient,  {under the effect of radial migration}.

\begin{figure}
\begin{center}
\includegraphics[width=1\hsize]{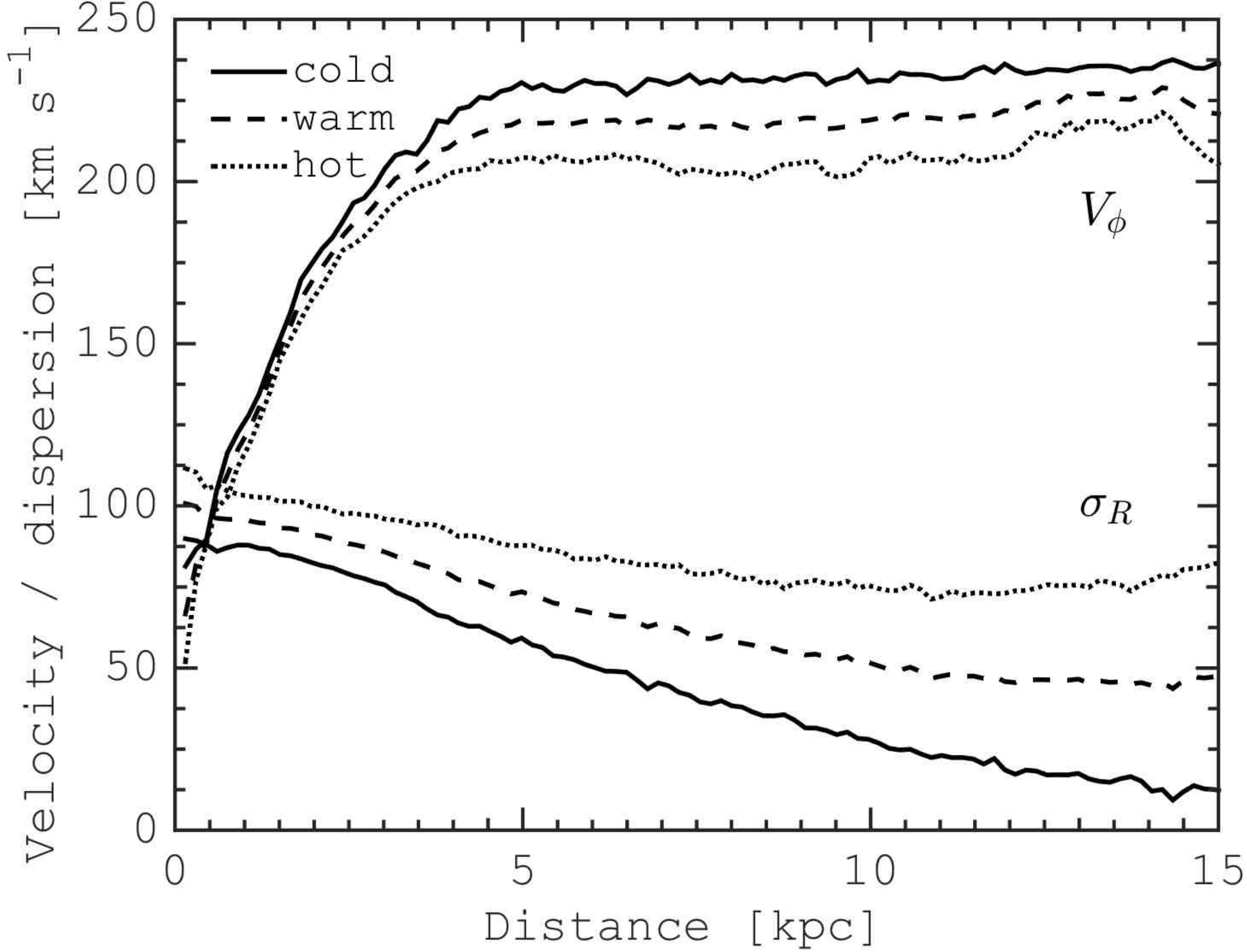}\caption{Initial profiles of the rotational velocity $V_{\phi}$ and the radial velocity dispersion $\sigma_{\rm R}$ for the hot~(solid lines), warm~(dashed lines) and cold~(dotted lines) disk populations.}\label{fig::fig1}
\end{center}
\end{figure}

Disk galaxies contain several populations of stars including bulge, thick and thin disks as well as stellar halo. 
Their chemical and kinematic characteristics provide information about their epochs of formation and the secular and dynamical processes that shaped them  {over time}~\citep{2002ARA&A..40..487F}. Stellar disk populations of different ages have a wide range of chemical and kinematical characteristics~\citep{2011ARA&A..49..301V}. In particular,  {in the disk of our Galaxy, the} velocity dispersion of stars depends  on their age and metallicity. Although the exact shape of the age-metallicity-velocity dispersion relation is  {still} debated, it is evident that younger stars  {tend to} have a  {higher} metallicity and smaller random  {motions} than older stars,  {which are on average more metal-poor and have higher random motions} ~\citep[see, e.g.,][]{1991A&A...245...57M,2007A&A...475..519H}. Hence metal-rich and metal-poor stars in a disk like that of our Galaxy  can be described as `kinematically cold' and `kinematically hot' components, respectively, and -- because of their different kinematics -- their contribution to spiral arms is expected to be different.

In this work, we study the  {differential response of stars in a composite disk galaxy -- made of kinematically cold and hot populations -- to spiral perturbations and the induced azimuthal} variations  {in the} mean metallicity  {of the disk}. We present a model in which stellar populations with different initial kinematics, but same radial density distribution, differentiate as  {a} spiral structure appears in the galactic disk. Here we want to make some testable predictions which can be checked particularly in the Milky Way disk with ongoing and future spectroscopic surveys~(e.g., APOGEE, Gaia, WEAVE). 

\section{Model}
We performed a single $N$-body simulation of  {the evolution of a composite} stellar disk,  {embedded} in a live dark matter halo.    {The disk in this simulation is made} of three co- {spatial} populations~(cold, warm and hot disks) with different velocity dispersions, but the same radial density distribution. Each disk component is represented by $2\cdot 10^6$~particles   {redistributed following a Miyamoto-Nagai density profile, with a characteristic scale length of }$3$~kpc  and a mass of $2.86\cdot10^{10}$~\Msun.  {The cold, warm and hot populations have characteristic scale-heights respectively equal to} $0.15$, $0.3$ and $0.6$~kpc. Our simulation also includes a live dark matter halo~($4\cdot10^6$ particles)  {whose density distribution follows a Plummer sphere, with}  a total mass of $4\cdot10^{11}\Msun$  {and} a radius of $21$~kpc. This approach is similar to recent studies by~\cite{2017MNRAS.469.1587D} and \cite{2017A&A...607L...4F}. The  {initial} kinematics   {-- circular velocity and dispersion profiles --} of these components is presented in Fig.~\ref{fig::fig1}. Initial  {conditions have} been generated using the iterative method described in~\cite{2009MNRAS.392..904R}. We used a parallel tree code by Khoperskov et al. (in prep) with the adoption of the standard opening angle $\theta=0.7$ and a gravitational softening parameter equal to $50$~pc. For the time integration, we used a leapfrog integrator with a fixed step size of $0.2$~Myr.

\begin{figure}
\begin{center}
\includegraphics[width=1\hsize]{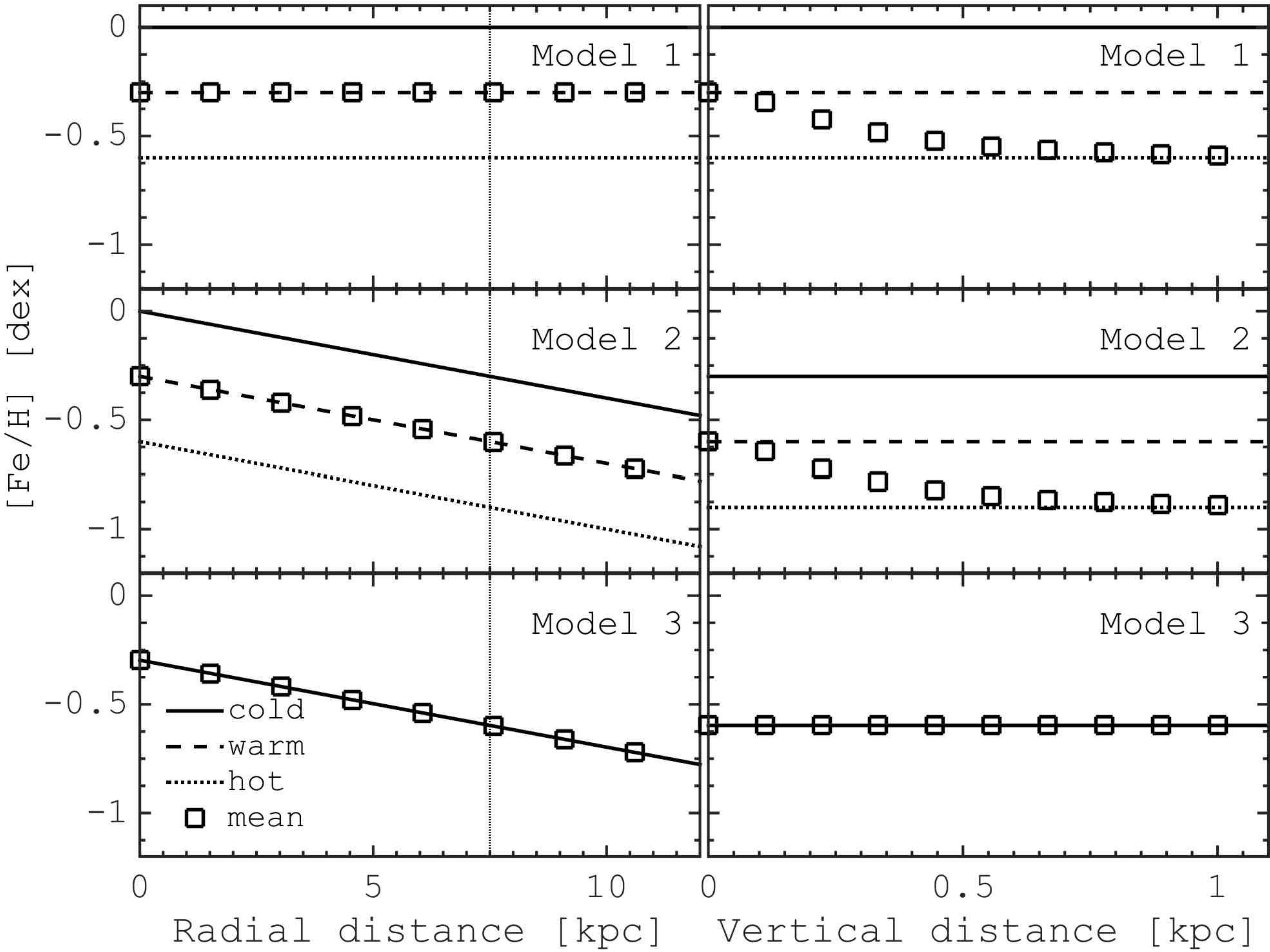}\caption{Initial metallicity profiles adopted for the different models: radial profiles in the equatorial plane~(left panels) and vertical profiles at $R=7.5$~kpc~(right panels).}\label{fig::fig0}
\end{center}
\end{figure}

\begin{figure*}
\begin{center}
\includegraphics[height=0.53\vsize]{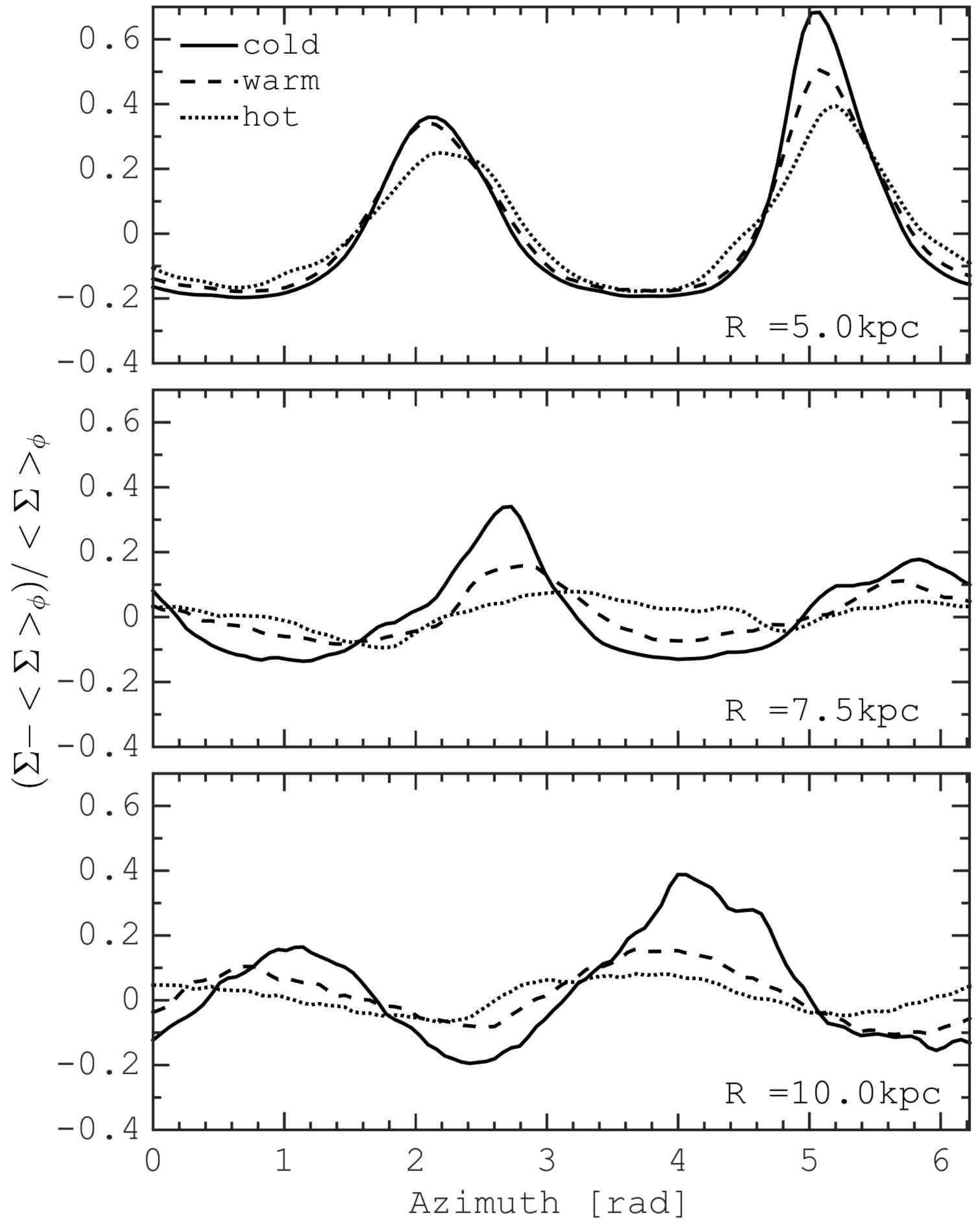}\includegraphics[height=0.53\vsize]{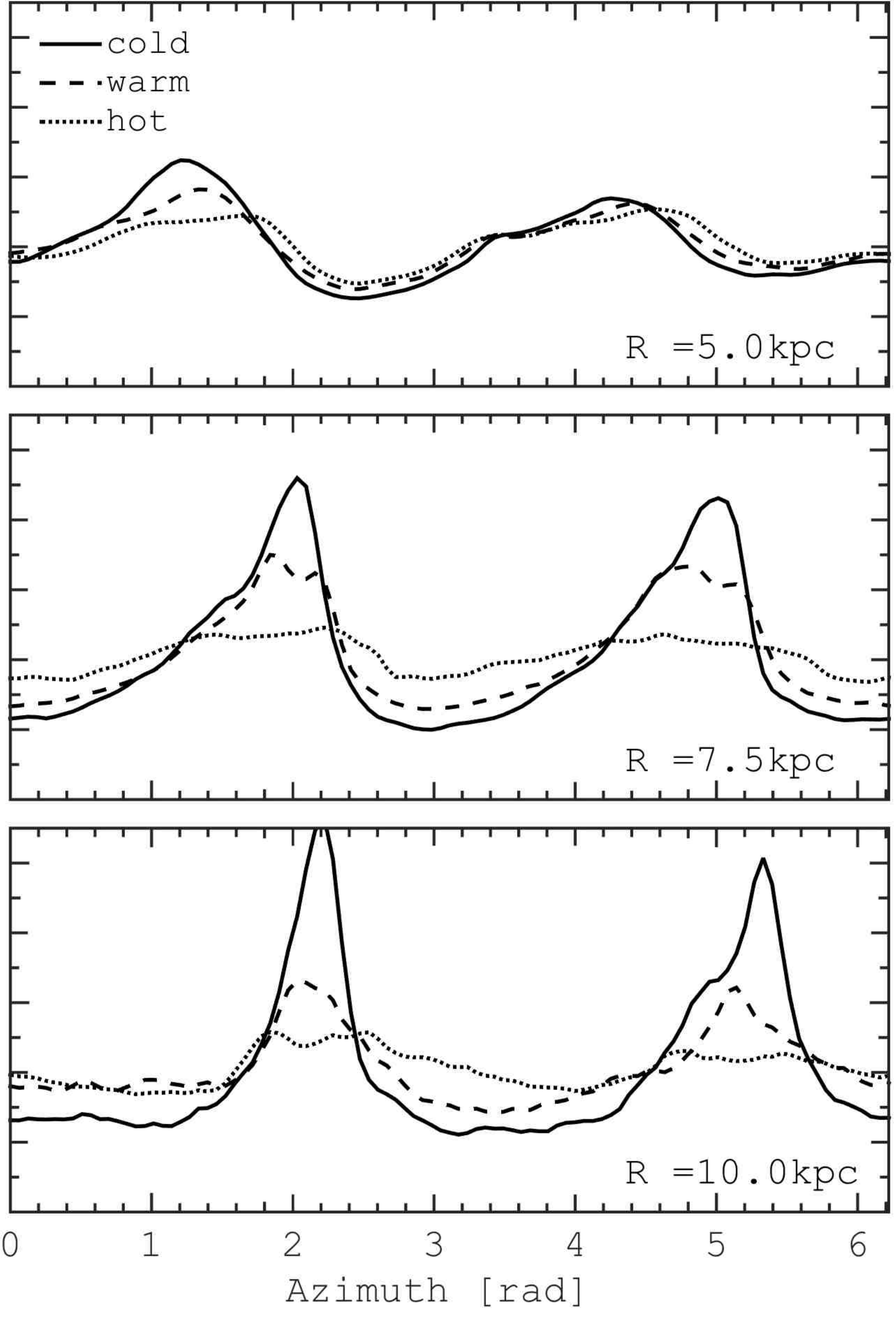}\includegraphics[height=0.53\vsize]{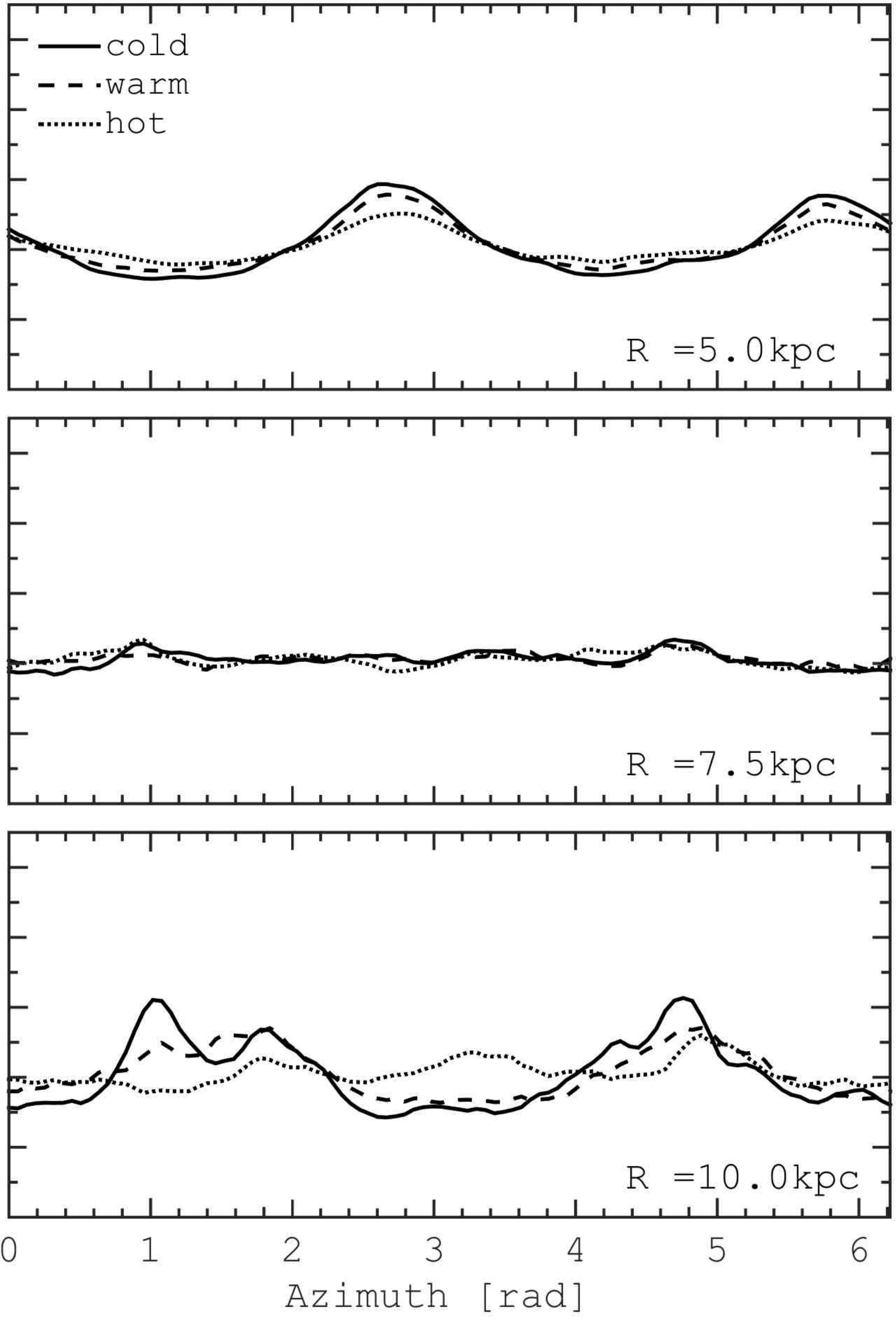}\caption{Azimuthal variations of the density perturbation for the cold~(solid), warm~(dashed) and hot~(dotted) disk components at different radii:~$R=5$~(top panels), $R=7.5$~(middle panels), and $R=10$~kpc~(bottom panels) and at different times: $t=0.3$ ~Gyr~(left column), $t=0.5$~Gyr~(middle column) and $t=1.3$~Gyr~(right column).}\label{fig::fig2}
\end{center}
\end{figure*}

\begin{figure}
\includegraphics[width=1\hsize]{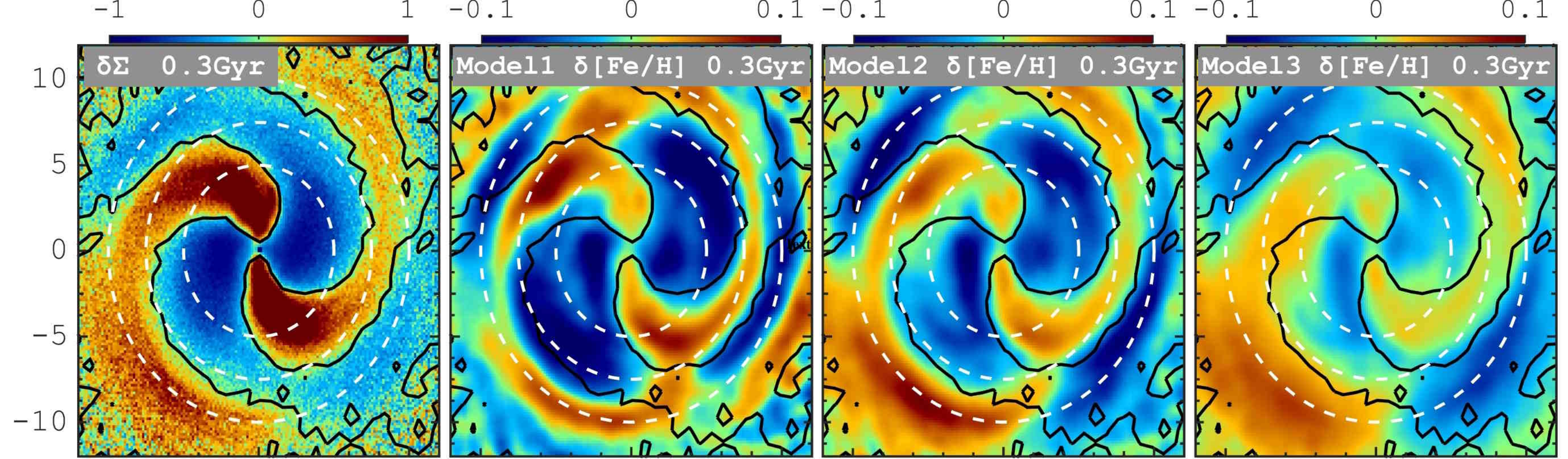}\\\includegraphics[width=1\hsize]{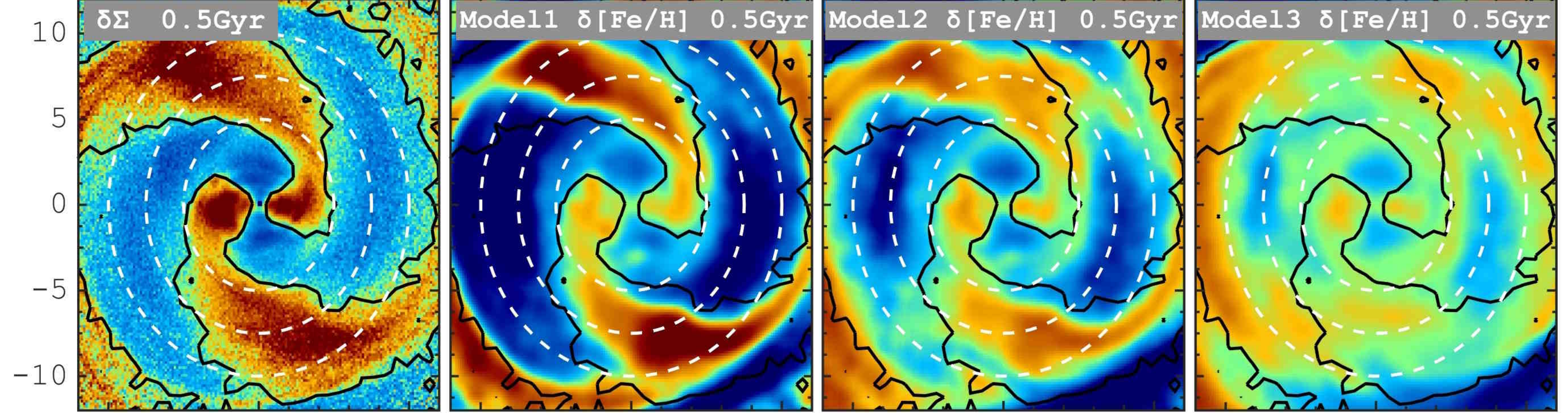}\\\includegraphics[width=1\hsize]{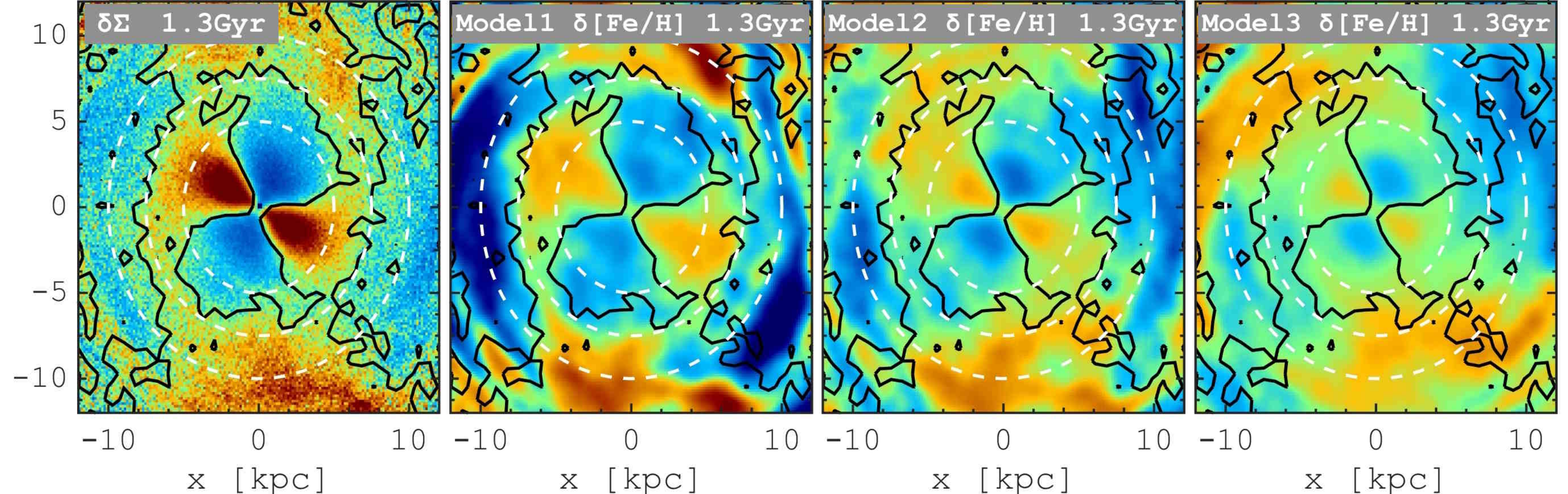}\caption{
Maps of the differential surface density~$\delta \Sigma$ (first column) and of the azimuthal metallicity variations~$\rm \delta[Fe/H]$ for Model~1, Model~2 and Model~3 (from the second to the fourth column), respectively. The different rows show the $\delta \Sigma$ and $ \rm \delta [Fe/H]$ maps at different times, corresponding to  $t=0.3$~Gyr~(top), $t=0.5$~Gyr~(middle) and $t=1.3$~Gyr~(bottom). In all panels, the solid black contour represents $\delta \Sigma=0$, and dashed circles show the radii~$R=5$, $R=7.5$ and $R=10$~kpc.}\label{fig::fig3}
\end{figure}

\begin{figure*}
\begin{center}
\includegraphics[height=0.56\vsize]{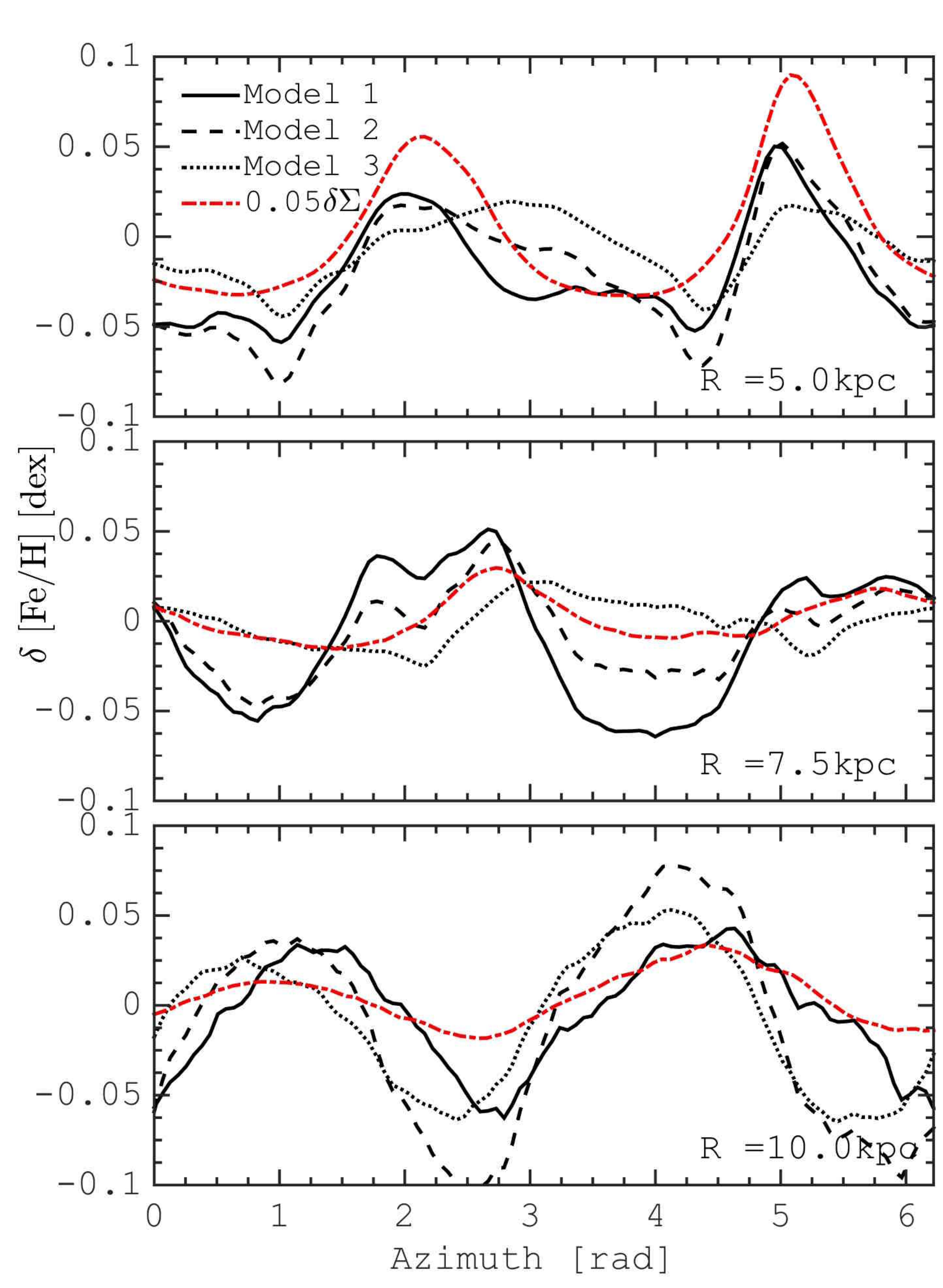}\includegraphics[height=0.56\vsize]{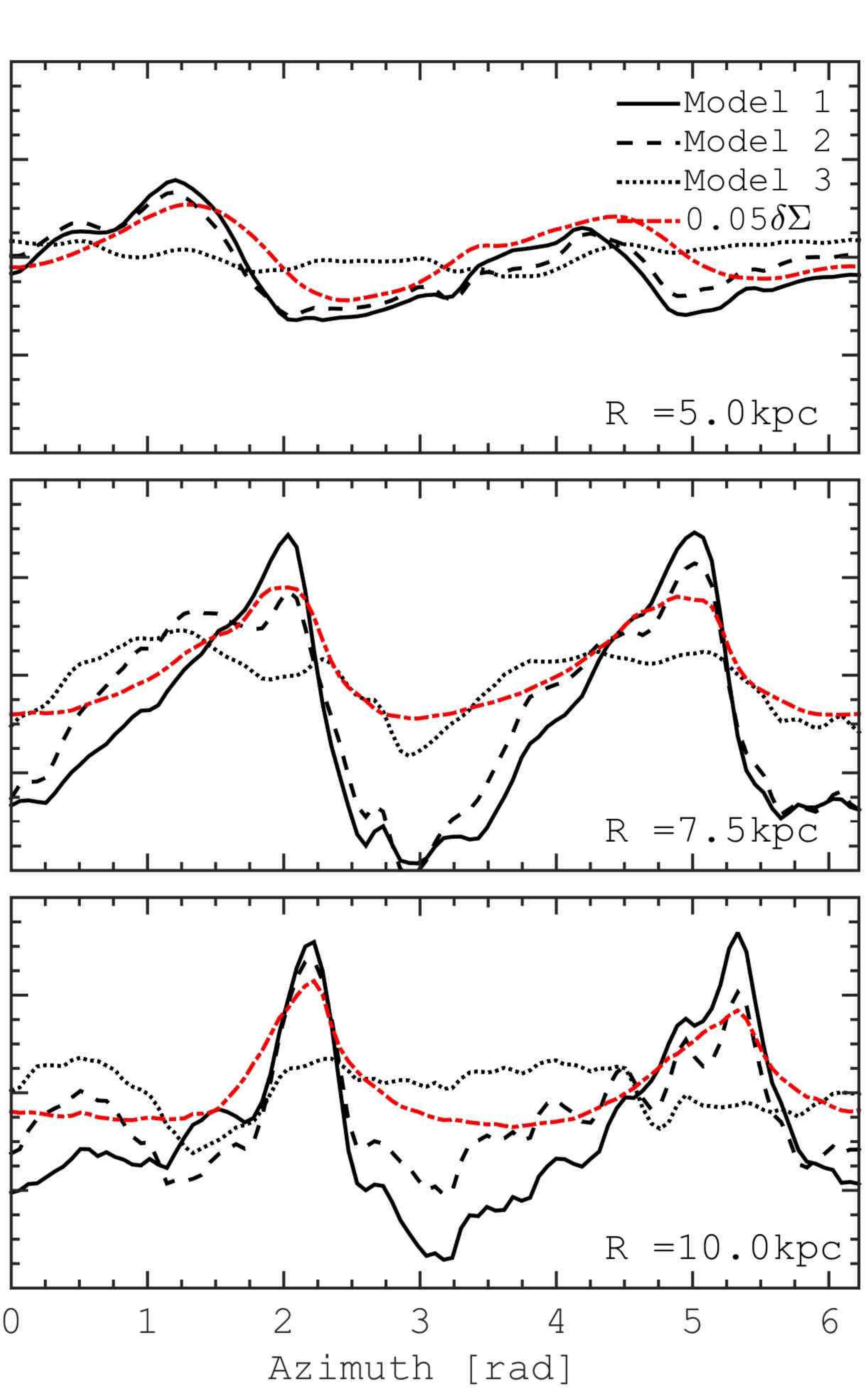}\includegraphics[height=0.56\vsize]{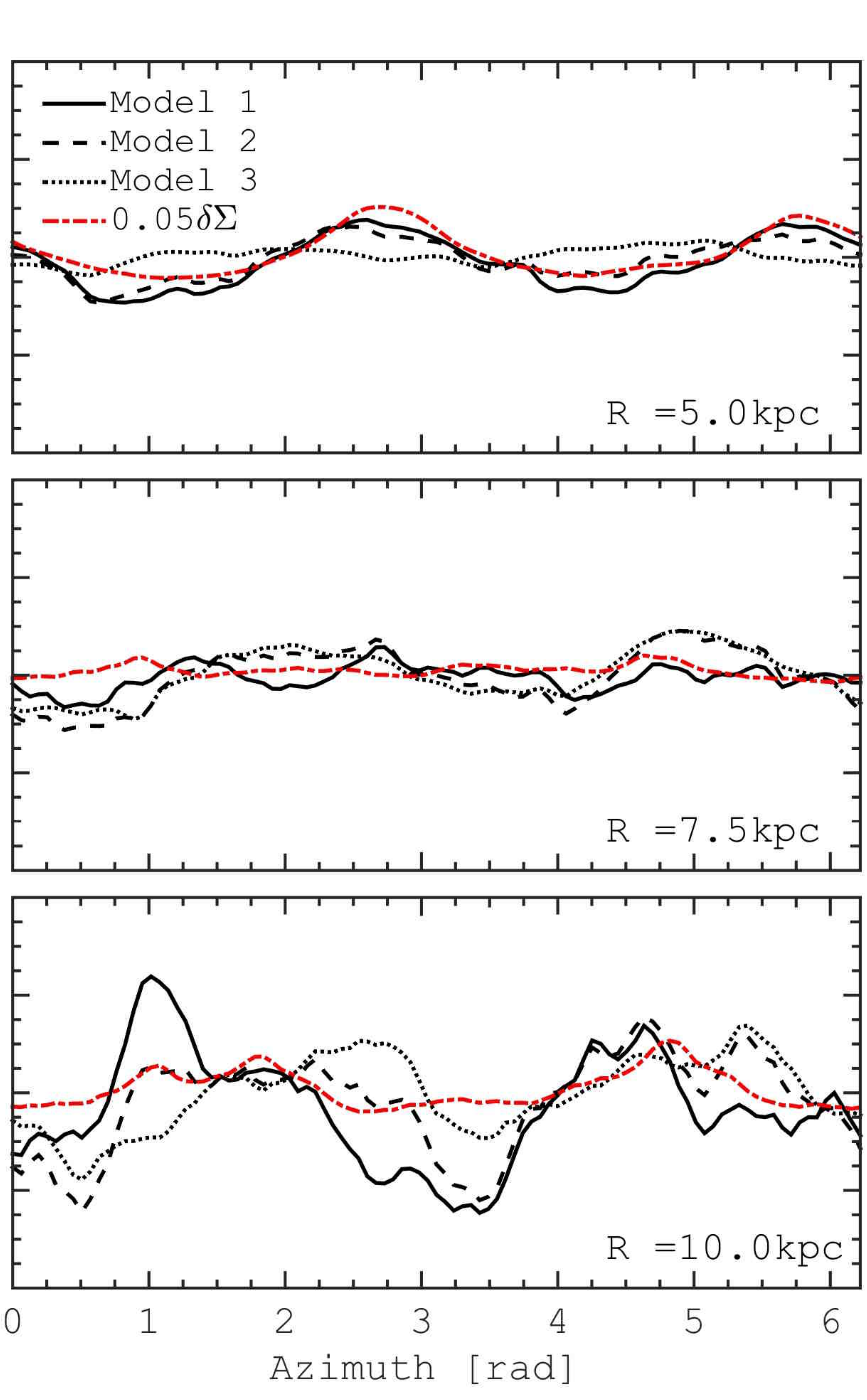}\caption{Azimuthal profiles of  metallicity variations~($\rm \delta[Fe/H]$) at different radii~($R=5$~(top panels), $R=7.5$~(middle panels), and $R=10$~kpc~(bottom panels)) for the different models. The profiles are shown at three times: $t=0.3$~Gyr~(left columns), $t=0.5$~Gyr~(middle column), and $t=1.3$~Gyr~(right column). The red lines represent the differential surface density ~($\delta \Sigma$), as a function of azimuth, at the same radius. Note that, to be easily compared with the metallicity variations, the values of ~$\delta \Sigma$ have been rescaled, see legend. }\label{fig::fig4}
\end{center}
\end{figure*}

 {Our simulation is dissipationless, and as a consequence we do not model star formation and chemical enrichment. However, we can trace the metallicity distribution and its temporal evolution, by flagging each particle in the disk with a corresponding metallicity. Depending on the way we assign metallicities to our disk particles, we can generate different models and quantify the strength of the azimuthal variations and how these latter are related to the initial distribution of metals in the disk. In particular, in this paper, we analyze three different models~(see Fig.~\ref{fig::fig0}).}
In Model 1 we assume that each disk population is characterized by a unique value of metallicity~($\rm [Fe/H]$),  {independent on the distance $R$ from the galaxy center, and} equal, respectively, to $-0.6$~dex, $-0.3$~dex and $0$~dex for the hot, warm and cold disk. No dependency on $z$, the height above the plane, is modeled, for none of the three disk components. \footnote{The choice of these values is in agreement with the mean characteristics of the stellar populations of the Milky Way disk~\citep[see, e.g.,][]{2012ApJ...753..148B}. We note, however, that the exact value of the metallicities of each population is not critical, the most important being their relative variations of metallicity}  {However, because the simulations  contains populations with initial different thickness, and these populations have different mean metallicities, a vertical gradient  naturally arises in the initial disk, when all populations are considered together.}
In Model 2  {we assume that each population has an initial radial metallicity gradient equal to $-0.04$~dex~kpc$^{-1}$ but different central (i.e. at $R=0$) values of the metallicity}.  {This model thus results initially in both a radial and a negative metallicity gradient, when the disk populations are analyzed all together.} 
 {Finally, in Model 3, the three populations have all the same radial gradient  ($-0.04$~dex~kpc$^{-1}$)  and the same central value of the metallicity, that is identical trends of the metallicity with $R$. No vertical metallicity gradient is assigned for any of the three populations. As a consequence of these assumptions,  Model 3 represents initially a disk with a radial metallicity gradient, but no vertical metallicity gradient.   }
 {We emphasize here that, }since we analyze the variations of the mean metallicity, we assume no metallicity scatter  {in any of the populations} . We checked, however, that an initial metallicity dispersion of $0.15$~dex does not lead to any differences in the results.

\section{Results}
We focus our analysis on two snapshots of the evolution at $0.3$, and $0.6$~Gyr, when a well-formed two-arm spiral structure is present in all three disk components. At later times the spiral structure decays, and we show a single snapshot at $1.3$~Gyr in order to underline that the signatures we found are related to the presence of spiral arms. In Fig.~\ref{fig::fig2} we show the relative perturbation of the stellar surface density where quasi-harmonic variations can be seen,  {with}  the cooler components showing the sharper density profiles. This is because in a co-moving rest frame stars of warmer components have a larger kinetic energy within the potential well of the spiral arms. Thus a warmer component is less sensitive to spiral perturbations than a cooler one, with stars from the latter spending a longer time in the spiral arm regions. Consequently, inside the spiral arms, the fraction of dynamically cold stars is  higher than that of warmer stellar components. In the inter-arm regions, we find the opposite, with  the warm and dynamically hot components contributing, for the majority, to these regions.   {Thus,}  stellar   {disk}  populations   {having initially different random motions}  are spatially differentiated in the presence of a spiral perturbation,   {similarly to what found}  for co-spatial populations in the Milky Way bar~\citep{2017MNRAS.469.1587D,2017A&A...607L...4F}.

We now consider possible variations in the metallicity distribution induced by this kinematic differentiation of stellar populations in the presence of  spiral patterns. In Fig.~\ref{fig::fig3} we show 2D maps of the differential surface density, $\rm \delta \Sigma= (\Sigma(R,\phi) - \langle \Sigma(R,\phi) \rangle_\phi)/\langle \Sigma(R,\phi) \rangle_\phi$, and of the azimuthal metallicity variations, $ \rm \delta [Fe/H]= [Fe/H](R,\phi) - \langle [Fe/H](R,\phi) \rangle_\phi$\footnote{In the definition of both quantities, the brackets $\langle  \rangle_\phi$ indicate the mean -- i.e. azimuthally averaged -- value of the quantity at a given radius.} for the three models~(see Fig.~\ref{fig::fig0}). As soon as the spirals start to develop, the spatial distribution of metals becomes asymmetric. We note that the metallicity patterns do not always trace precisely the density variations. In our Model~1~(with no initial radial metallicity gradient for none of the disk components, but with an initial global vertical gradient) we find a very prominent spiral-like periodical variations of metallicity with a substantial contamination of metal-rich stars~(cooler components) in spirals and of metal-poor stars~(hotter components) in the inter-arm regions. These azimuthal variations of metallicity remain strong as long as the spirals arms are present in the disk ($t<1$~Gyr).  {We emphasize here that, since in this model stars belonging to the same disk population have all the same metallicity, the azimuthal variations in the metal distribution found are $only$ a manifestation of the different response of these disk populations to the bar and spiral perturbation.} Less contrasting but similar results are obtained for Model 2 where the only difference with respect to Model~1 is that, in  each disk component, stars have a metallicity which depends on their distance $R$ from the galaxy center. For Model 3, where identical radial metallicity profiles are imposed for all three populations,  {and, as a consequence, at all radii, stars in the cold, warm and hot populations have all the same metallicity}, the azimuthal variations in the metal distribution are $only$ a manifestation of the radial migration -- by churning and blurring -- of stars in the disk induced by the bar and spiral arms~\citep[][]{2008ApJ...684L..79R}. Stars initially in the inner disk, and which have -- by construction --  the highest metallicities, can migrate outwards via spiral arms, and vice versa stars from the outer disk can migrate inwards, this giving rise to the metallicity patterns found.  {Note that in Models 2 and 3 there is a flattening with time of the radial gradient,  which} is a well known result of radial migration of stars via resonance scattering~\citep[see, e.g.,][]{2009MNRAS.398..591S,2012A&A...540A..56P}. 

To analyze the variations of metallicity more quantitatively, in Fig.~\ref{fig::fig4} we show the azimuthal profiles of the metallicity variations, $\rm \delta [Fe/H]$. The patterns show that azimuthal variations in the metallicity in all models appear as soon as stellar asymmetries start to develop. At early times the amplitude of the metallicity variations is comparable in all our models, but in the inner disk region, it is more pronounced in Model 1 without radial gradient. Since we observe the formation of a bar in the innermost part~($R<5$~kpc), the impact of the spiral structure can be better appreciated at $R=7.5$~kpc, where the amplitude of metallicity variations is $\approx 0.1$, $0.08$ and $0.05$~dex for Models 1, 2 and 3 respectively.  At later stages of evolution when the spiral patterns decay~($t>1$~Gyr), for all models the metallicity distribution becomes more complex and shallow. In this case, systematic azimuthal variations of metallicity can be found only in the inner bar-region~($R<5$~kpc). Some small-scale structures appear at the outskirts where inhomogeneities are related to the radial migration of stars~\citep{2012A&A...548A.126M,2013A&A...553A.102D}. 

In our Model~1, the metal-rich peaks coexist with the maxima of the density perturbations~(Fig.~\ref{fig::fig4}), where the contamination of kinematically cool (metal-rich) population is maximal.  For Model~3 metal-rich peaks are mostly on the trailing side of the spiral, as it was suggested by~\citet{2016MNRAS.460L..94G}.  In this regard, we find a possible qualitative difference between purely kinematically driven metallicity patterns~(Model~1) and the azimuthal variations shaped by  radial migration~(Model~3). In reality, both effects (kinematic differentiation and radial migration) should possibly be taken into account in the interpretation of the metallicity patterns found in external galaxies and in the Milky Way disk.

\begin{figure}
\begin{center}
\includegraphics[width=1\hsize]{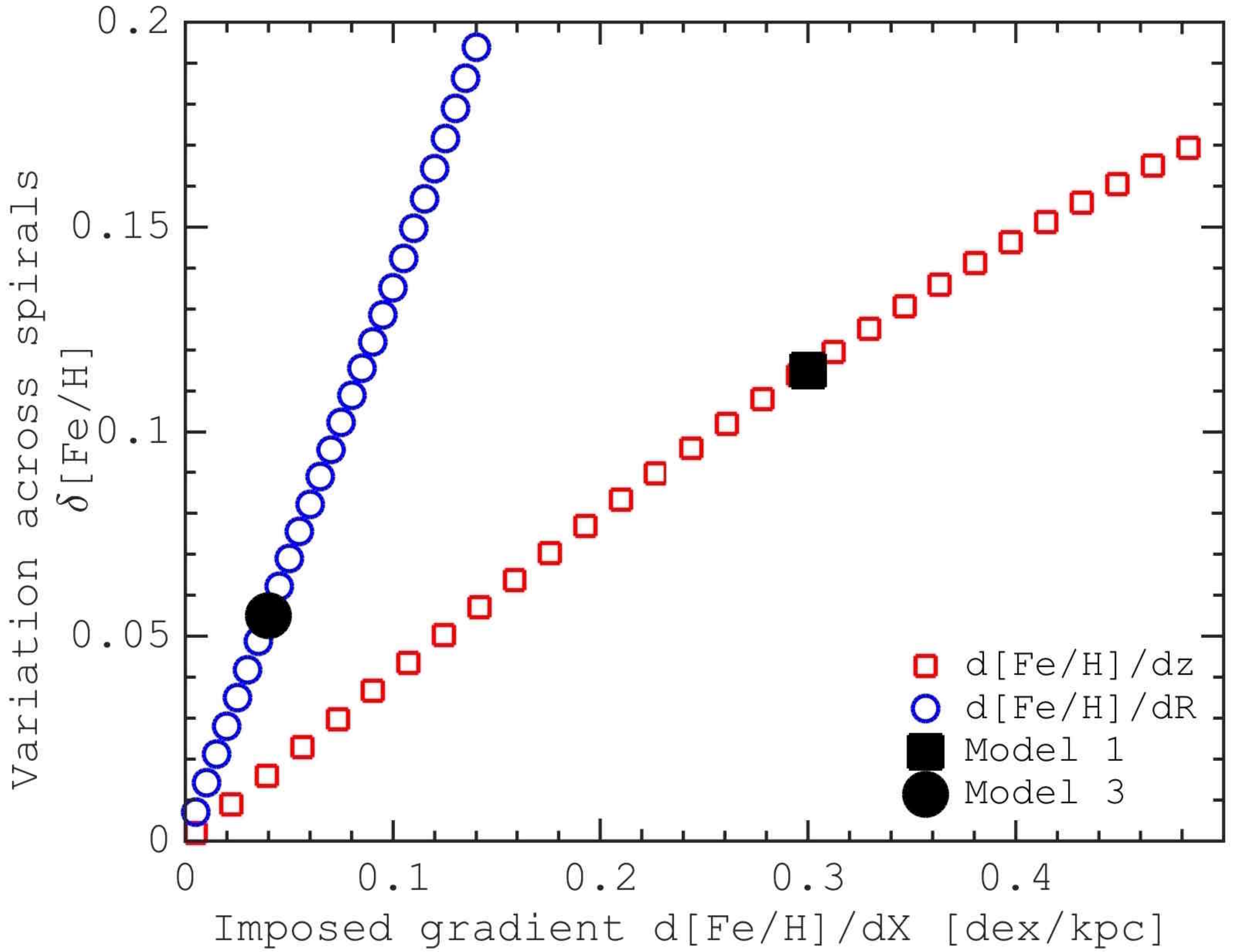}\caption{Amplitude of the azimuthal metallicity variations versus the initial vertical~(red squares) and radial~(blue circles) metallicity gradients, for a single snapshot at $0.3$~Gyr at $R=7.5$~kpc. The values corresponding to Model 1 and 3 are shown, respectively, with a filled square and a filled dot.}\label{fig::fig5}
\end{center}
\end{figure}

We can also compare the effect of initial radial and vertical gradients on the amplitude of the azimuthal variations in the metal distribution. To do this, we vary the initial metallicity values for the different components in Model~1 -- thus effectively changing the initial vertical metallicity gradient in the model -- and the strength of the radial gradient value in Model~3. In Fig.~\ref{fig::fig5} we plot the amplitude of azimuthal metallicity variation as a function of the initial vertical and radial gradients. We show that the vertical gradient should be by a factor of a few larger than in a radial direction to produce the same azimuthal variation of metallicity. In other words the azimuthal variations are more sensitive to the radial gradient. Fig.~\ref{fig::fig5} also shows a qualitative agreement with \cite{2016MNRAS.460L..94G}, where the azimuthal gradient is more pronounced with steeper radial~(or vertical) gradient.

\section{Conclusions}\label{sec::conclusions}

 {By means of a high-resolution $N$-body simulation}, we have studied the large-scale variations of the stellar metallicity in the disk of a spiral galaxy  {made of} co-spatial populations with different initial velocity dispersions. We show, as expected, that  {spiral patterns} are more sharply defined for dynamically colder populations and are  {smoother} for dynamically hotter ones.   {Because of this different response, } significant azimuthal variations in the distribution of metals in a galaxy disk are observed. Thus, pre-existing radial metallicity gradients are not a necessary condition for the existence of azimuthal metallicity variations, that can indeed be produced also in disks with initially null radial gradients, but vertical ones. Although the exact shape of the age-metallicity-velocity dispersion relation of stars in disk galaxies is debated, it is evident that younger stars have a lower metallicity and smaller random velocity component than the older ones. Thus we expect that the mechanism described in this paper is also able to produce azimuthal age variations in the disk. The existence and strength of azimuthal variations in the Milky Way and the influence of the bar and spiral arms on the distribution of metals are still open questions, but we will soon be able to investigate them with data from Gaia and follow-up spectroscopic surveys. It will then be interesting to check if these variations  can be explained by the effect described here.

\begin{acknowledgements}
We thank the anonymous referee for their comments and Ivan Minchev for stimulating discussions. This work was granted access to the HPC resources of CINES under the allocation 2017-040507 (PI : P. Di Matteo) made by GENCI.  This work has been supported by the ANR (Agence Nationale de la Recherche) through the MOD4Gaia project (ANR-15-CE31-0007, P.I.: P. Di Matteo) and by the RFBR grant~(16-32-60043).  
\end{acknowledgements}

\bibliographystyle{aa}
\bibliography{references}

\end{document}